DESIGN 2014

# Study of human accessibility: physical tests versus numerical simulation

M. Delangle, E. Poirson and J.F. Petiot
LUNAM - IRCCyN - Ecole Centrale de Nantes, 1, rue de la Noe, 44321
*Nantes*

*Keywords : Population model, Numerical simulation, Comparative analysis*

## 1. Introduction

In the field of universal design, taking into account the physical characteristics of the target population during the design process is essential. However, human variability causes difficulty to meet the requirements of comfort and security of a certain percentage of the user population. That's why human factors engineering allows to optimize the interactions with the user systems, considering the human body variability of population. Physical ergonomics allows to study interactions among humans and other elements of a system, to create an environment that is well-suited to a user's physical requirements. Many tools and practices are in common use today that assist in basic assessments of accommodation, and indicate to the engineer how to design for the variability in body dimensions (or anthropometry), capability and age, in the target user population [Roebuck, 1995].Commonly used approaches are the "population models" (where a group of individuals representing the target-user population interacts with a real prototype) and the use of anthropometric data where body dimensions are used to fit the relevant product dimensions.

To avoid building expensive physical mock-up, the designer can use anthropometrics database (e.g. ANSUR [Gordon et al., 1989], NHANES [for Disease Control and Prevention, 1994]) generally used to configure sizes of a product, to improve perception (e.g., comfort, safety, etc.) across a population to maximize the satisfaction of the greatest number of users. Thus, several methods based on these anthropometric data [Committee, 2004] are used to design adapted products and assess the degree of accommodation. Existing database may be chosen as the reference population and use directly ("boundary mannequins"...) or through regression techniques ("population model", hybrid approaches"...) [Garneau and Parkinson, 2007]-[Holler, 2010]-[Reed and Flannagan, 2000]. For example, "2D boundary mannequin" approach can used database, where relevant dimensions are taken directly from an anthropometric database to create manikins at the xth percentiles of the measure of interest, to represent a specified accommodation [Garneau and Parkinson, 2010]. Defining the mannequins so that they correctly represent the future users is frequently simplified into the use of a standard setup based on a few percentile values, or a combination of predefined key anthropometric variables [Hanson and Högberg, 2008]. In reality, a xth percentile may show poor correlation between body size and may not include physiological situations occurring frequently as tall people with relatively short arms or small people with relatively long legs [Marshall et al., 2002]. Although numerical methods are faster and less expensive than building prototypes, the only use of these anthropometric databases [Moroney and Smith, 1972], often old, are not always representative of the target user population. In fact, they often consist of specifics surveys (military...), and typically provide only very limited information concerning children and people who are older and disabled [Marshall et al., 2002]. Moreover, this methods are generally used in univariate case study (to determine the appropriate allocation of adjustability to achieve a desired accommodation level), where most problems are multidimensional. That is why, although the use of digital data allows to avoid to perform experimental tests and building prototypes, methods based on this principle still pose questions about the ease of use and reliability compared to reality. So, we propose to perform an accessibility study compare results obtained from simple numerical models (using raw anthropometric data) with those obtained experimentally.

## 2. Comparative study

In this study, four "simply" tests are perfomed to evaluate accessibility based on physical capacity (not including preference).

### 2.1 Methodology

This paper considers two ways to assess accessibility in which experimentation and database approach are applied in accessibility evaluation (Fig.1). We focus on the accessibility of the upper body. A population of various people is selected. They are asked to participate to two different tests of accessibility (reached envelope): discret test A and continuous test B. In the same time, some of there physical characteristics are measured to create an anthropometric database. These database aims to used as input of the two numerical simulations to determine the theoretical reached envelope. The two experiments proposed, each time realized physically and numerically, aims to highlight the differences of experimental results in comparison of real accessibility, and so verify if the numerical simulation of the upper body reach envelope might be considered as realistic.

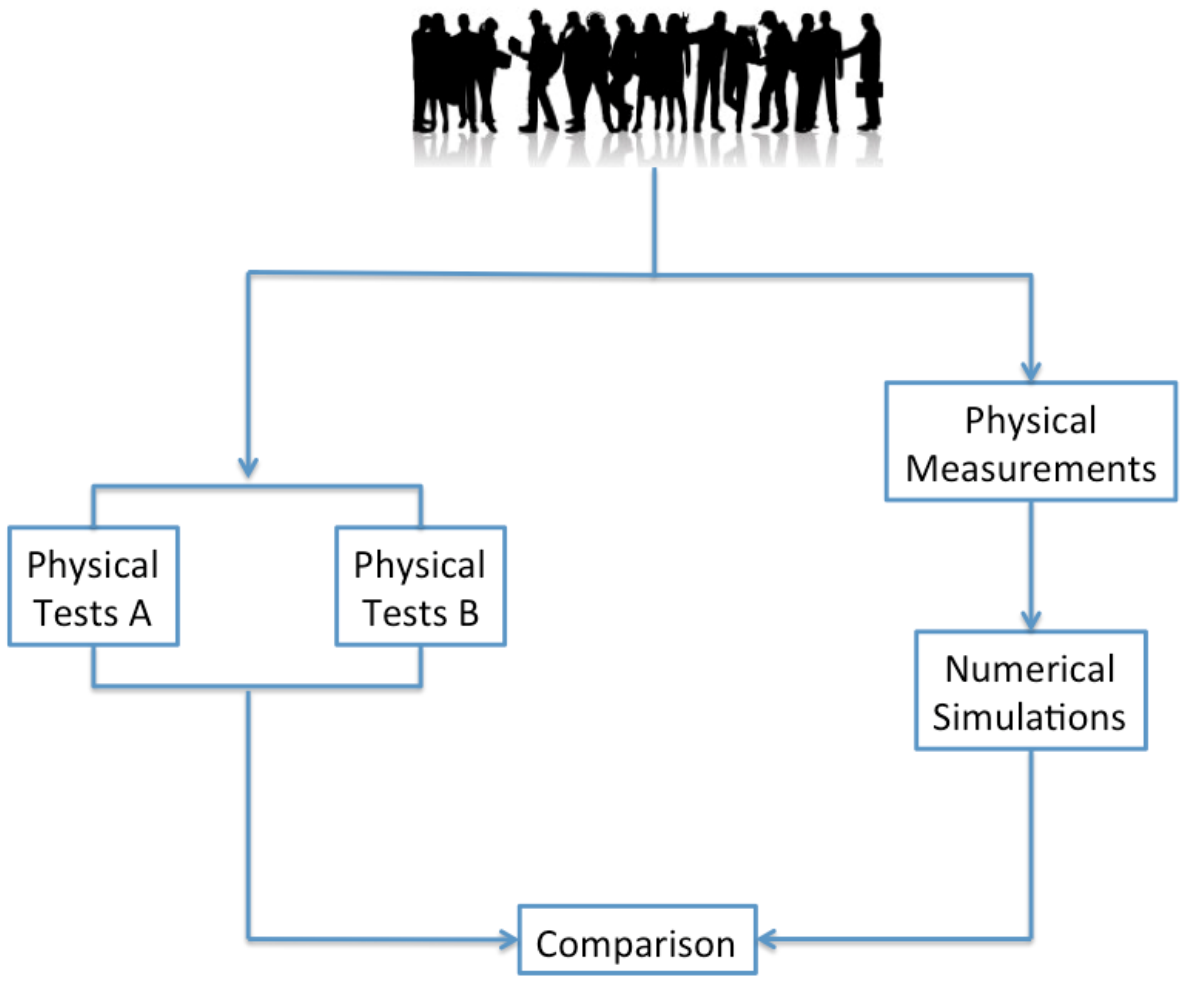

**Figure 1. Synopsis of the study.**

### 2.2 Subjects and anthropometric measurement

The experiments were conducted with 40 adult volunteers, all of being French students or teachers of the Ecole Centrale de Nantes. In order to create a database of measurements, four anthropometric characteristics were measured for each of them (Fig.2), namely the stature, the shoulder height, the shoulder width and the arm length (Table.1). Twenty five males and fifteen females were sampled in the study, covering a wide spectrum of physical characteristics, from 1482 mm for the smallest stature, to 1930 mm for the highest.

**Table 1. Anthropometric characteristics measured for each participants, with values in mm.**

| Individuals | Gender | Stature | Shoulder height | Shoulder width | Arm length |
|---|---|---|---|---|---|
| 1 | M | 1735 | 1485 | 470 | 750 |
| 2 | F | 1705 | 1450 | 430 | 730 |
| 3 | F | 1482 | 1240 | 420 | 640 |
| ⋮ | ⋮ | ⋮ | ⋮ | ⋮ | ⋮ |
| 40 | M | 1715 | 1420 | 460 | 710 |

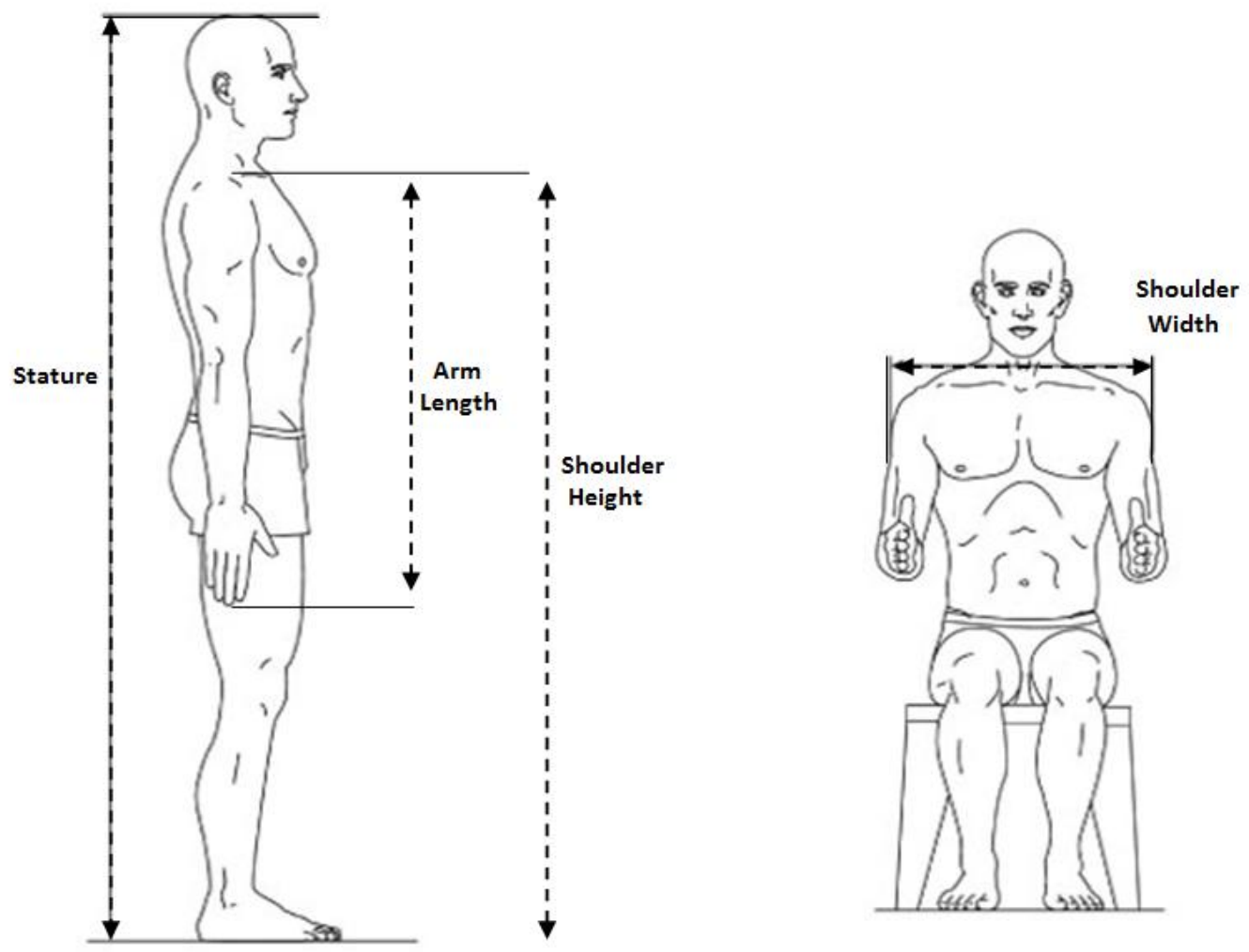

**Figure 2. Anthropometrics characteristics measured.**

## 2.3 The A Test

Experiment A allows to simulate realistic arm accessibility situations through interaction with physical points (switches).

### 2.3.1 Description and physical realization

**Principle**: *Accessibility to switches located in front of an individual and positioned on a vertical plate (effect of the body position while reaching).* A plate is positioned on a vertical wall with the aid of two adjustable vertical axes for precise positioning of the bottom plate at shoulder height He (Fig.3). Reach measurements are made relatively to a body reference point (shoulder joint) and to a measurement apparatus point (bottom of the plate). The subject is positioned in the center of the plate with the feet fixed regarding the floor (position sensor is positioned under the heels of the subject indicating if the feet are off the ground or not). For each individual, the reached switches are identified and noted in a table in order to draw the reach envelop. A total of 84 switches can be reached on the plate, which constituted 168 measurement points (two positions for each switch). With a view to ease of notation and understanding, participants are asked to touch the switches by color strips (with his left or right arm according the side), giving 12 black , 32 white, 44 green, 46 blue and 34 red reachable points. The device is designed to fit at shoulder height for a wide range of human physical characteristics (designed from 2.5th and 97.5th percentile for women and man stature of ANSUR database), allowing to perform the tests for a large user population. Because most anthropometric data presented in databases represent nude body measurements and to permit reliable comparison with database approach, experiments are performed with light clothing (nude dimension and light clothing being regarded as synonymous for practical purposes). For both tests A1 et A2, the task demands can be defined as "reach switches with at least one finger touches the switch and pushing it".

**Test A1** : The subject keeps the feet fixed to the floor and the body must stay fixed (spinal column axis) relative to the vertical axis of the center of the plate, and reaches switches one arm at a time.
**Test A2** : The subject keeps the feet fixed to the floor but is allowed to twist (changing the position of the upper body) to reach the switches, one arm at a time.

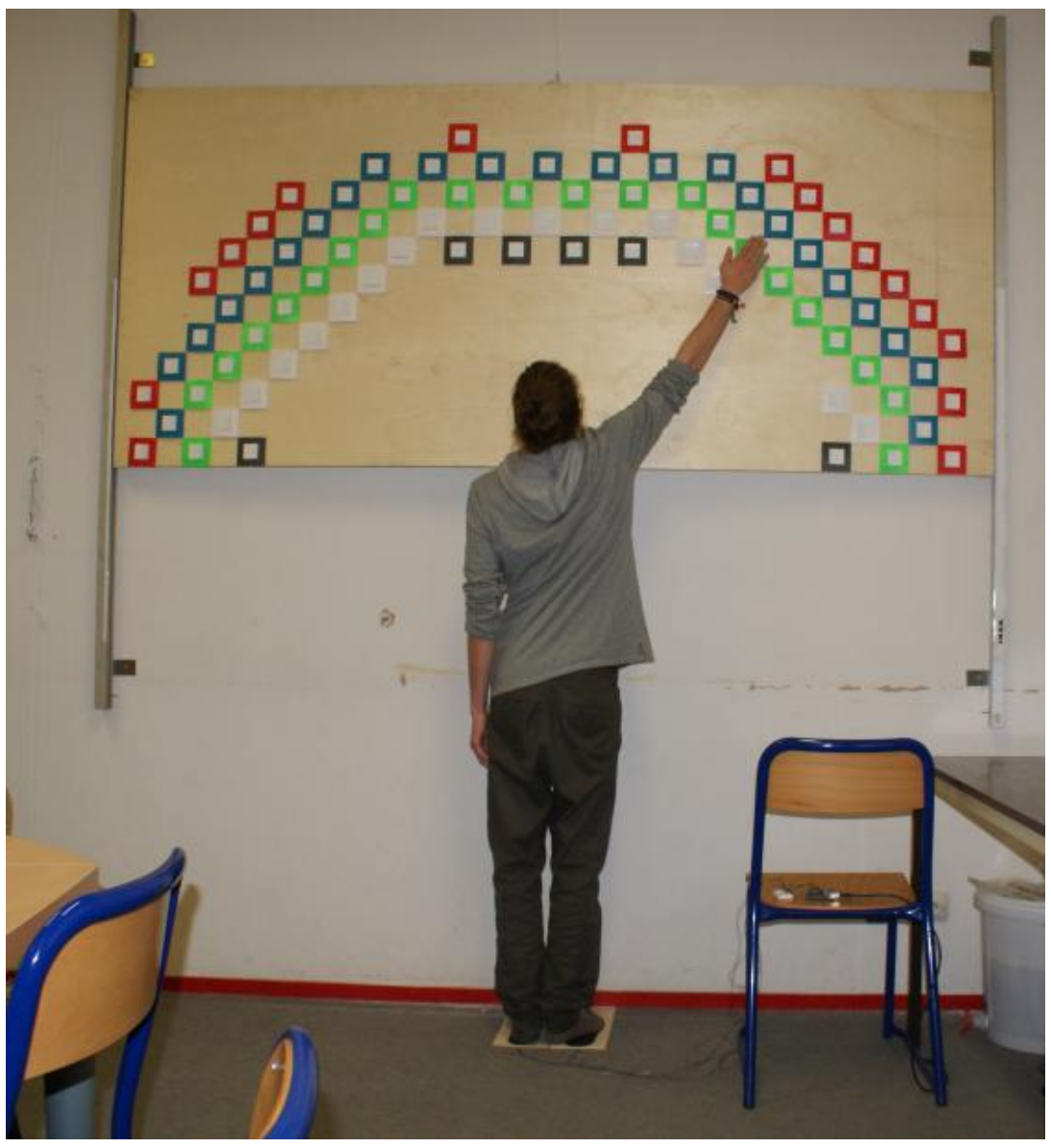

Figure 3. **Discret experimental test A1 (the subject switches reaching switches on the right of the plate with his right arm).**

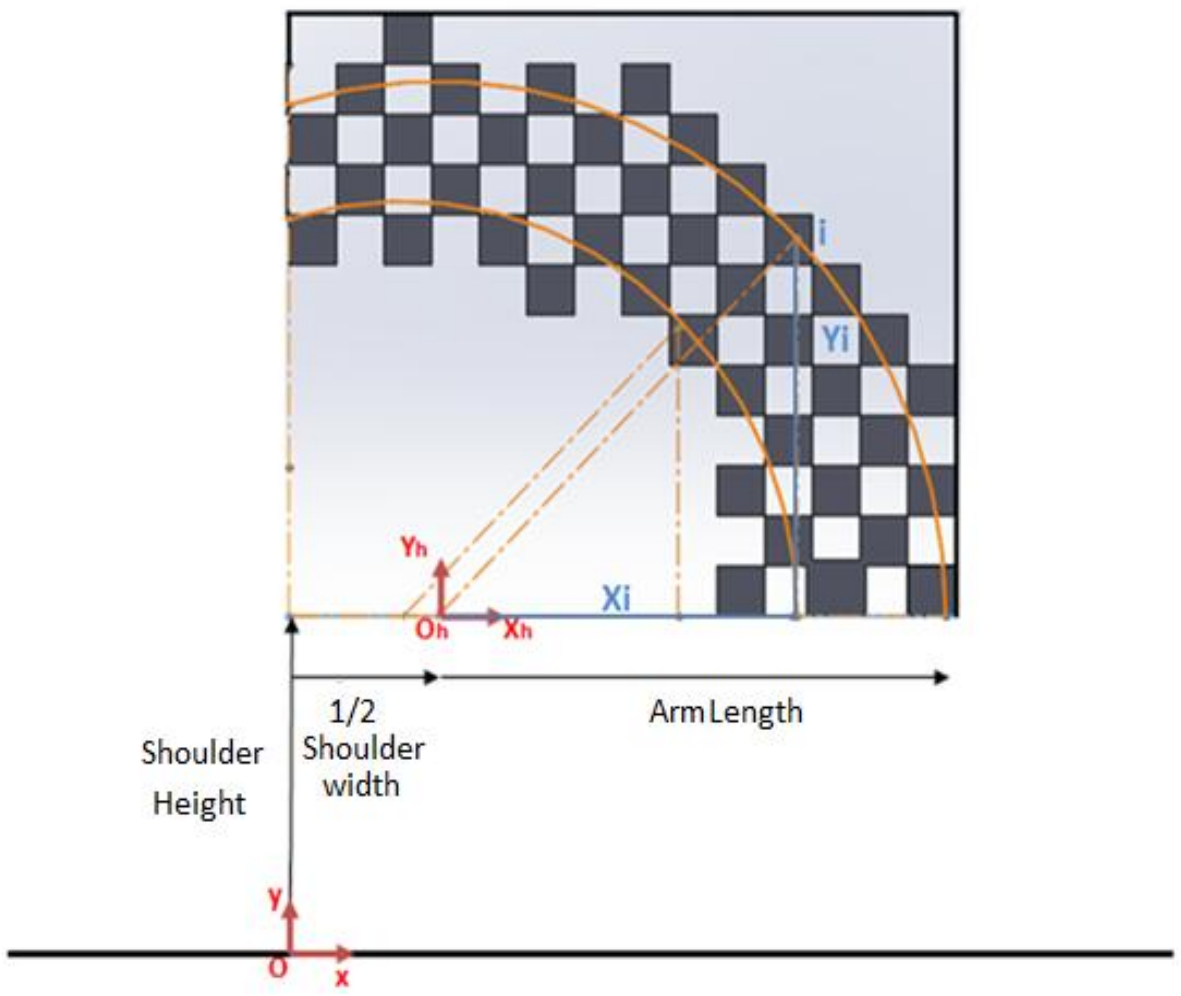

**Figure 4. Definition of accessibility of is from anthropometric data.**

**2.3.2 Numerical simulations (Tests A)**
**Reached switches calculation** : *The objective is to compute the reached switches according to the anthropometric characteristics recorded during experimentations (experimental database).* The reach envelop is defined by a circle arc, with the arm length Larm as radius and the position of the shoulder Oh as rotation point (Fig.4). So, knowing the coordinates of the switches on the plate, the theoretically reach points computed from arm length and shoulder width can be defined. Reach envelope is defined by assuming that the arm makes a perfectly circular arc with the shoulder as point of rotation. Moreover, as the database usually provide data for only one arm, we use the value of the right arm and admit that both are perfectly symmetrical. Knowing the switches coordinates on the plate and the anthropometrics characteristics of individuals, a programme is implemented (using Matlab R2012b) allowing to automatically determinate which switches are theoretically reached by the arms. The realisation of experimental tests A1, A2 and numerical calculations provides, for each participant:
_ Switches with the fixed body
_ Switches achieved with the upper body flexible
_ Switches that should have been achieved according to the model used

**2.4 Experimental tests B**
**2.4.1 Description and physical realization**
**Objective** *: Marking of the boundaries of arm reach of an individual in standing posture.*
The participant stays in front of an erasable marking device, positioned on a vertical wall on which he must draw his reached envelope (Fig.5). Then, a photo of the drawing is taken and treated thanks to a Matlab program to accurately determine the area of the envelope (Fig.6). As tests A, reach envelope of tests B are constructed considering one-handed operation with fixed foot, and movements or not of the torso.

*Note* : reach design dimensions or envelopes for design use should be constructed considering grasp requirements which may affect the functional reach envelope. Fingertip touch resulting in the largest reach dimensions appropriate for touch controls, envelopes are marked to define the task as "finger touch" function (one finger touches an object without holding it) in order to avoid much as possible grasp effect (reducing the reach envelope).

**Test B1** : The subject keeps the feet fixed to the floor and the body fixed (spinal column axis) relative to the vertical axis of the center of the plate, and marks his reach envelop with both arms simultaneously.

**Test B2** : The subject keeps the feet fixed to the floor but is allowed to twist (changing the position of the upper body) to marks his reach envelop one arm at a time.

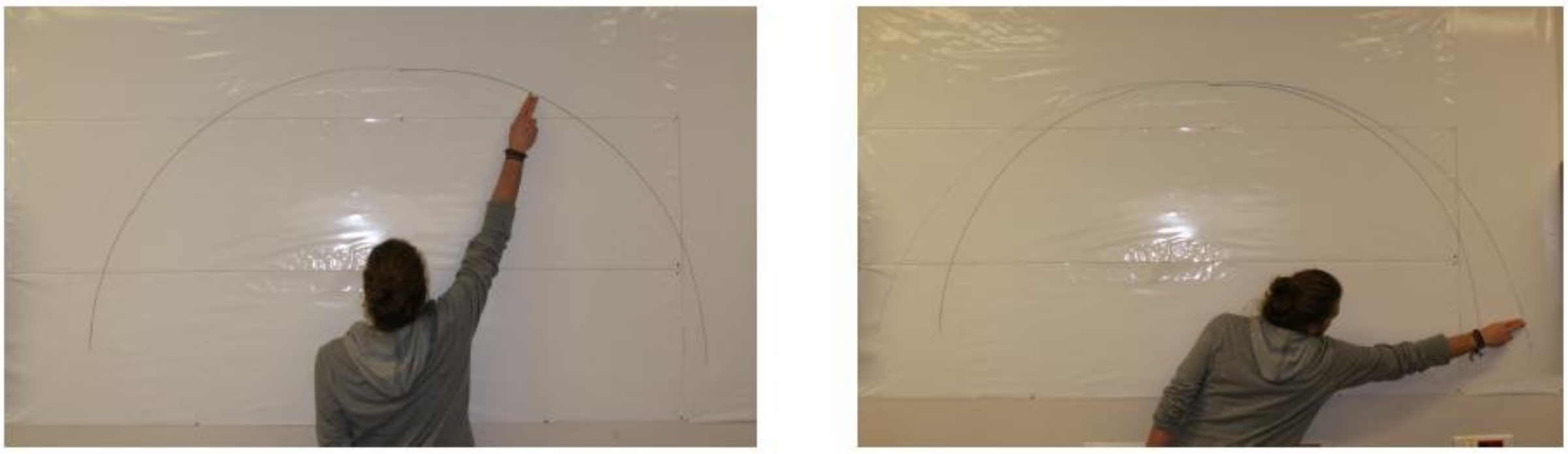

**Figure 5. Continuous experimental test B1 (left) with the feet and the body fixed, and test B2 (right) with the feet fixed and the upper body in movement.**

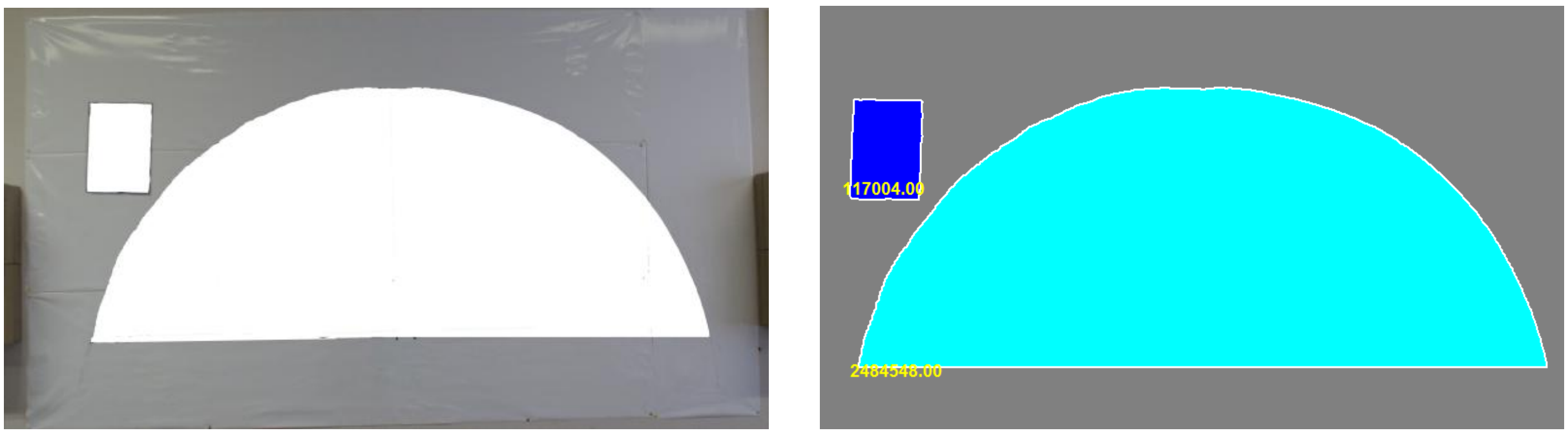

**Figure 6: Image processing using matlab. Bright space represents the reach envelope area drawn by the participant. The dark area (rectangle) represents the paper sheet used as repository to calculate areas from pixels to square meters.**

### 2.4.2 Numerical simulations (Tests B)

The objective is to determine the reach envelope area of a subject based on his anthropometric characteristics, such as might be found in a database. That is, reach envelope area for each participant (defined by envelopes Fig.7), is calculate using the arm length [iB] and the shoulder width [ih] (all other quantities being calculate from these values). Reach envelopes should be constructed considering the shoulder (arm rotation point) as reference point in the envelope calculation to determine the different arc circle areas. So, the total area reached on the wall by the arms of the individual is numerically simulated by Eq.1, where both arm lengths are considered as the same. Area A4 is a common part for both arm; so to avoid redundancy in the calculations, this area is not taken into account in the definition of the total envelope.

$$A_{totale} = 2.(A_1 + A_2 + A_3) = 2.\int_i^B \sqrt{B^2 - x^2}\mathrm{d}x + 2.\int_i^h \sqrt{B^2 - x^2}\mathrm{d}x + h.y_{Hi} \quad (1)$$

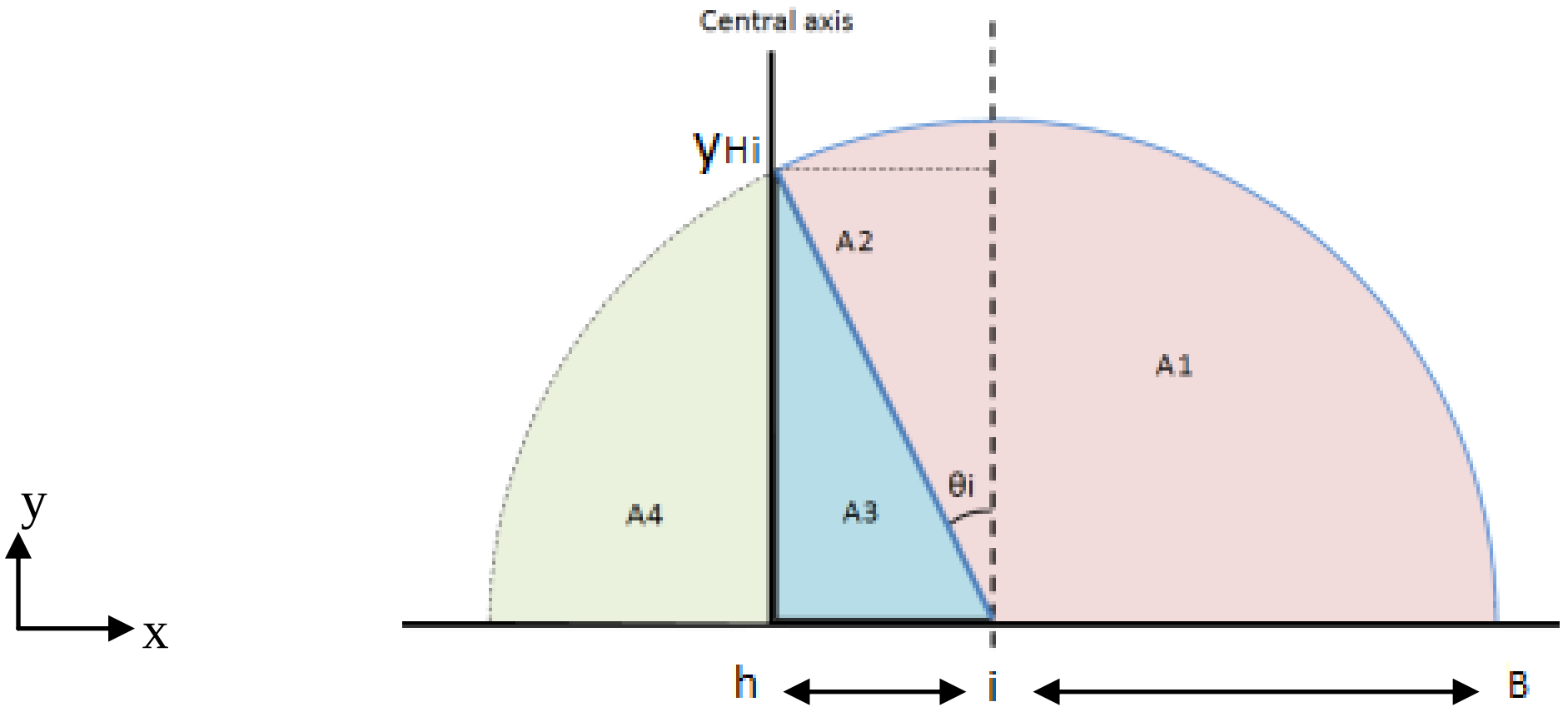

**Figure 7. Calculation of reach envelope areas.**

## 3. Results and discussion

Each participant was asked to perform the four experiments. Results are collected, providing for each of them the experimental results obtained with fixed body, with flexible upper body and from calculations, in the case of discret tests (Table 2) and continuous tests (Table 3). Figure 9 and 8 represent the results of tests A1, A2, B1 and B2 measured from experiments, and the numerical calculations plotted against stature for the 31-member sample, with regression line.

**Table 2. Numbers of reached switches for each participants, measured from experiments A1, A2 and numerically calculated. ΔA1 and ΔA2 are respectively the differences of reached switches between the calculations and experiments.**

| Individuals | 1 | 2 | 3 | 4 | ... | 31 |
|---|---|---|---|---|---|---|
| Reached switches for A1 test | 79 | 67 | 30 | 86 | ... | 71 |
| Reached switches for A2 test | 107 | 102 | 52 | 109 | ... | 97 |
| Reached switches numerically calculated | 81 | 59 | 30 | 91 | ... | 60 |
| $\Delta A1$ | 2 | -8 | 0 | 5 | ... | -11 |
| $\Delta A2$ | -26 | -43 | -22 | -18 | ... | -37 |

**Table 3. Reach areas for each participants, measured from experiments B1, B2 and numerically calculated. ΔB1 and ΔB2 are respectively the differences of areas between the calculations and experiments.**

| Individuals | 1 | 2 | 3 | 4 | ... | 31 |
|---|---|---|---|---|---|---|
| Reach areas for B1 test | 1.13 | 1,06 | 0,85 | 1,26 | ... | 1,22 |
| Reach areas for B2 test | 1.28 | 1,29 | 1,01 | 1,65 | ... | 1,42 |
| Reached areas numerically calculated | 1,23 | 1,15 | 0,91 | 1,31 | ... | 1,11 |
| $\Delta B1$ | 0,10 | 0,08 | 0,05 | 0,04 | ... | -0,10 |
| $\Delta B2$ | -0,05 | -0,15 | -0,10 | -0,34 | ... | -0,30 |

The aim of the experiments was to prove :
- the influence of the stretching of the body on the reach
-the quality and the limits of the numerical simulation thanks to the real experiment.

Regarding table 2, we notice firstly that the number of switches activated might be very different: from 30 to 105. It is very dependent of the stature: Pearson coefficient is 0.83 against 0.78 with the length of arms. The average standard deviations of the distributions A, A1 and A2 are respectively equal to 21, 23 and 24. We compare the significance of the stretching of the body. Results for A2 are obviously superior to A1 but the range is hight: an average increasing of 40% of switches. The second point is about the validation of our model. If we compare A1 to A from simulation, we find an average difference of 8 switches. Knowing that the switches are organized clothes one from the other, this difference is acceptable. But in the case of A2, this same average is 33. As it might be expected, the model used is not adapted to body with a possible stretching. On the Fig.7, it can be seen a vertical translation between the linear regressions obtained from experimental data A1, A2 and calculated data A.

The calculate data B and experimental data obtained through tests B1 and B2, plotted with respect to stature, are shown Fig.8. The average standard deviations of the distributions B, B1 and B2 are respectively equal to 0:12, 0:15 and 20:2. This dispersion of the calculated data might be explained by the differences of anthropometry of subjects, accentuated by the fact that the male and female data are processed simultaneously, explaining the cloud of points obtained. For example, for a same height, the arm lengths can vary greatly from one individual to another, directly affecting the reach envelope, which is calculated from the value of arm length and shoulder width. The results obtained experimentally by tests B1 are quite close to those obtained theoretically, with a mean difference $\Delta B1 = (B_{calculated} - B1_{experimental}) = 0,031$. So, the reach areas measured from experiment B1 might be considered as acceptable in comparison of reality. However, as for the test A2, values for B2 are widely superior in comparison to those calculated with a mean difference of $\Delta B2 = (B_{calculated} - B2_{experimental}) = -0,239$.

In general, it is observed that the values obtained experimentally are higher compared with numerical calculations from anthropometric data of individuals passing the tests. The results show the wide range of results which these various methods all of which are in common use can provide to this extremely simple accessibility assessment. The tests A1 and B1 have allowed to show that in a fixed position, the results obtained by numerical method are quite close to reality. However, results from tests A2 and B2 have shown numerical calculations varying greatly from reality. This is due to the fact that in this situation, accessibility is measured when the body is in motion and engaged in a physical activity. Indeed, the numerical model (which uses raw numerical data) being based on static anthropometry, does not take into account the functional anthropometry induced by dynamic behaviour of the human body (stretch of muscle, body motion ...) intrinsic to real situations.. These finding are consistent with other studies ([Kroemer et al., 1990]) involving more complex problem showing that reach limits are clearly dependent on the task, motion, and function to be accomplished by the reach action.

So, our study has demonstrated that (1) Stretching the body is far from negligible be in computing accessibility, and (2) the static model is not representative of the stretched body (functional anthropometry) but simply the right body (static anthropometry). It is also interesting to highlight the impact of the functionality of the task at hand. Indeed, testsA involving physical touch points, experiment shows that in this case, even if experiments are implemented to minimize this behavior, participants tend to unconsciously want to move his upper body to reach switches. The differences in behavior highlights the impact of real-life situations (notions of objective and performance) through the physical points (switches) on the results. So, this shows that task demands (touching the switches) might affect reach characteristics and measures.

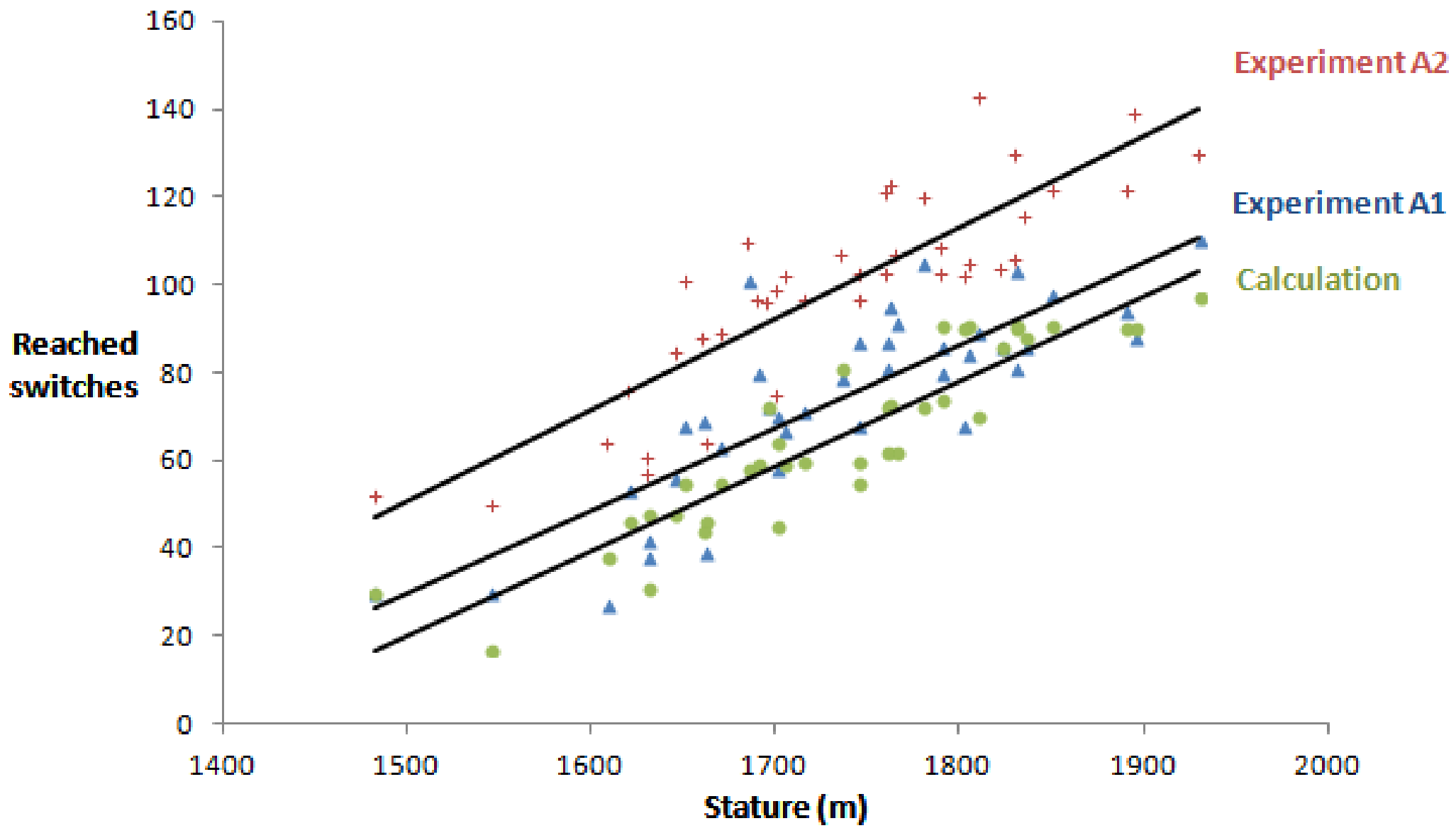

**Figure 8. Results for A1 and A2 measured from experiments. and from numerical calculations, with regression line.**

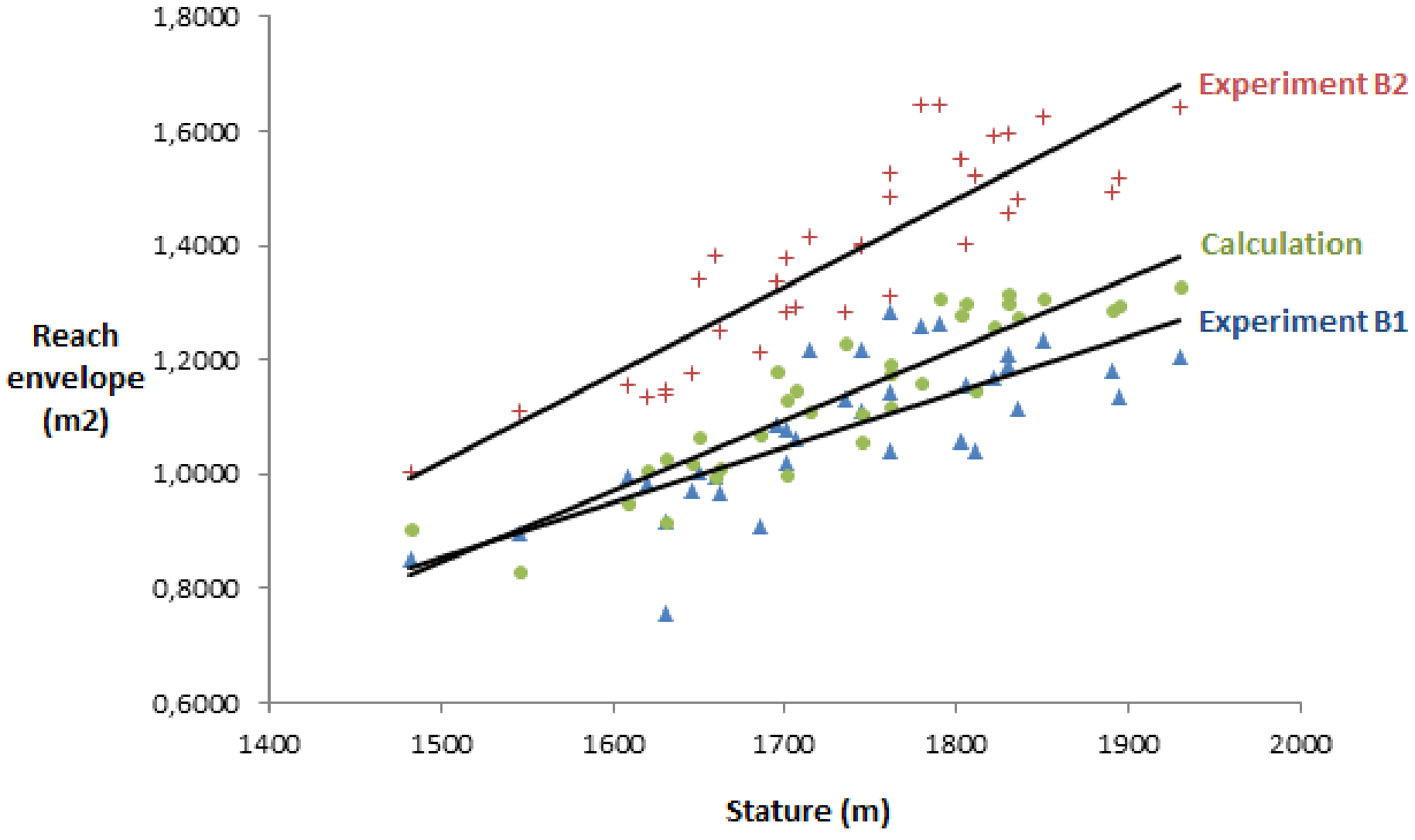

**Figure 9. Results for B1 and B2 measured from experiments and from numerical calculations, with regression line.**

Initial study trials have shown the limits of the numerical simulation (based on raw anthropometric data) thanks to the real experiment (influence of the stretching of the body on the reach). The aim for further work is to use a kinematic model to predict dynamic reach, and compare results with those in this study. Moreover, study will be performed with more subjects in order to have a significant number of participants to create a new statistical model of accessibility according to the anthropometry. An other perspective would be to explore models of body fatigue regarding the required task. In other words, although stretching the muscles can increase the achievement envelope, if a such movement is harmful to humans, these kind of "extreme" skills have no interests in point of view of ergonomic.